\documentclass[prd,a4paper,preprint]{revtex4}
\usepackage{graphicx}

\begin{document}

\title{A mass formula for light mesons from a potential model}
\author{Fabian \surname{Brau}}
\thanks{FNRS Postdoctoral Researcher}
\email[E-mail: ]{fabian.brau@umh.ac.be}
\author{Claude \surname{Semay}}
\thanks{FNRS Research Associate}
\email[E-mail: ]{claude.semay@umh.ac.be}
\affiliation{Universit\'{e} de Mons-Hainaut, Place du Parc 20,
B-7000 Mons, BELGIQUE}
\date{\today}

\begin{abstract}
The quark dynamics inside light mesons, except pseudoscalar ones, can be
quite well described by a spinless Salpeter equation supplemented by a
Cornell interaction (possibly partly vector, partly scalar). A mass
formula for these mesons can then be obtained by computing analytical
approximations of the eigenvalues of the equation. We show that such a
formula can be derived by combining the results of two methods: the
dominantly orbital state description and the Bohr-Sommerfeld
quantization approach. The predictions of the mass formula are compared
with accurate solutions of the spinless Salpeter equation computed with
a Lagrange-mesh calculation method.
\end{abstract}

\maketitle

\section{Introduction}
\label{sec:intro}

Semirelativistic potential models have been proved extremely successful
for the description of light mesons (mesons containing $u$, $d$, or $s$
quarks). The main characteristics of the spectra of these mesons, except
pseudoscalar ones, can be obtained with a spinless Salpeter equation
supplemented with the Cornell interaction (a Coulomb-like potential plus
a linear confinement) \cite{fulc94,brau98}.

Numerous techniques have been developed in order to solve numerically
with a great accuracy the semirelativistic equation. Nevertheless, it is
always interesting to work with analytical results. Several attempts to
obtain some mass formulae for hadrons were already performed. Some
approaches rely on fundamental QCD properties \cite{sing82,morp90}, but
they are limited to the study of ground states of hadrons. In other
works, the hadron masses are given as a function of some quantum
numbers. They are based, for instance, on shifted large-$N$ expansion
($N$ is the number of spatial dimensions) of the Schr\"{o}dinger
equation \cite{pagn86}, a spectrum generating algebra \cite{iach91}, or
a completely phenomenological point of view \cite{sema95}. We will adopt
here a different point of view by assuming that a semirelativistic
potential model allows a good description of the main features of meson
spectra.

Recently, a new method to tackle this problem was developed: the
dominantly orbital state (DOS) description, in which the orbitally
excited states are obtained as a classical result while the radially
excited states are treated semiclassically
\cite{goeb90,olss97,silv98,brau00a}. A second method is the
Bohr-Sommerfeld quantization (BSQ) approach, with which precise
information can be obtained on the asymptotical behaviors of observables
as a function of the quantum numbers \cite{brau00b}. We show here that a
quite well accurate mass formula for light mesons, as a function of
quantum numbers and parameters of a QCD inspired potential, can be
obtained by combining the results of these two approaches. The idea is
to calculate analytical approximate solutions of the equation assumed to
govern the quark dynamics inside a meson. There is yet some
uncertainties about the Lorentz structure of the interquark interaction.
In this work, we will assume that the confinement potential is partly
scalar and partly vector. A related work using a WKB approach was
performed in Ref.~\cite{cea82}, but the Coulomb-like potential and the
Lorentz structure of the confinement interaction were not taken into
account.

In order to test the validity of the mass formula, we compare its
predictions with accurate numerical solutions of the spinless Salpeter
equation. These last ones are computed with a Lagrange-mesh calculation
method \cite{sema01}. This technique is here modified in order to handle
semirelativistic equations with mixed scalar-vector potentials.

In Sec.~\ref{sec:model}, the model Hamiltonian is presented with the two
methods previously developed to compute some analytical solutions. The
mass formula is established in the case of symmetric and asymmetric
mesons, with our without a constant term in the potential. In
Sec.~\ref{sec:disc}, the mass formula is compared with accurate
numerical solutions of the spinless Salpeter equation. Some concluding
remarks are given in Sec.~\ref{sec:concl}.

\section{The model}
\label{sec:model}

\subsection{Model Hamiltonian}
\label{ssec:modh}

Within the framework of a semirelativistic potential model, it is
possible to describe the main characteristics of the
spectra of light mesons \cite{fulc94,brau98}. If the spinless Salpeter
equation is chosen, instead of the Schr\"{o}dinger equation, the
quark-antiquark Hamiltonian is given by
(we use the natural units $\hbar=c=1$)
\begin{equation}
\label{hsv}
H=\sqrt{\vec {p\,}^2+\left( m_1+\alpha_1 S(r) \right)^2}
+ \sqrt{\vec {p\,}^2+\left( m_2+\alpha_2 S(r) \right)^2}
+ V(r),
\end{equation}
where $V(r)$ and $S(r)$ are respectively the vector and scalar
interactions \cite{goeb90}, and where $\vec p$ is the relative momentum
between the quark and the antiquark. The vector $\vec p$ is the
conjugate variable of the inter-distance $\vec r$. As usual, we assume
that the isospin symmetry is not broken, that is to say that the $u$ and
$d$ quarks have the same mass
(in the following, these two quarks will
be named by the symbol $n$).
The parameters $\alpha_1$ and $\alpha_2$
indicate how the scalar potential $S(r)$ is shared among the two masses
$m_1$ and $m_2$. A natural choice, used in this work, is to take
\begin{equation}
\label{alpha2}
\alpha_1 = \frac{m_2}{m_1+m_2} \quad \text{and}
\quad \alpha_2 = \frac{m_1}{m_1+m_2}.
\end{equation}

It is generally admitted that the short range part of the interquark
potential is dominated by the one-gluon exchange process, which gives
rise to a Coulomb term of vector type. The long range part is dominated
by a confinement that lattice calculations predict linear in the
interquark distance. As its Lorentz structure is not precisely known, we
suppose here that the confinement is partly scalar and partly vector, as
in
Ref.~\cite{olss97}. The importance of each one is reflected through a
parameter $f$ whose value is 0 for a pure vector, and 1 for a pure
scalar. Consequently, the potentials considered here are given by
\begin{equation}
\label{s}
S(r)=f\,a\,r
\end{equation}
in which $a$ is the usual string tension, whose value should be
around 0.2 GeV$^2$, and
\begin{equation}
\label{v}
V(r)=(1-f)\,a\,r-\frac{\kappa}{r}
\end{equation}
in which $\kappa$ is proportional to the strong coupling constant
$\alpha_s$. A reasonable value of $\kappa$ should be in the range 0.1 to
0.6.

It is worth noting that it is not possible to describe the pseudoscalar
mesons with a so simple potential. Spin contributions are very large in
this sector as well as flavor mixing effects. An interaction stemming
from instanton effects, which is not considered here, could explain the
properties of these mesons \cite{brau98,blas90}. Consequently, the
pseudoscalar mesons cannot be described by our model.

\subsection{Semi-classical method}
\label{ssec:semi}

Approximate analytical solutions of Hamiltonian~(\ref{hsv}) with
potentials~(\ref{s})-(\ref{v}) can be obtained within the DOS approach.
The idea of the model is to make a classical approximation by
considering uniquely the classical circular orbits (lowest energy
states with given total orbital angular momentum $J$), defined by $r$
constant, and thus $\dot{r}=0$. The radial excitations, numbered with
the quantum number $\nu$, are
calculated by making a harmonic approximation around the previous
classical orbits. A detailed description of this method is given in
Refs.~\cite{goeb90,olss97,silv98,brau00a}. We just recall here the main
results. In the case of a symmetric meson, $m_1=m_2=m$, the square meson
mass $M^2$ is given by \cite{silv98}
\begin{equation}
\label{m2sc}
M^2=a\,A(f)\, J+B(f)\, m\, \sqrt{a\, J} +C(f)\,m^2+a\, D(f) \kappa
+a\ E(f) (2\nu +1) + O(J^{-1/2}).
\end{equation}
The coefficients $A$, $B$, $C$, $D$, and $E$ are given by
\begin{eqnarray}
\label{abcde}
A(f)&=&\frac{y^2}{4}\left[t+3(1-f)\right]^2, \nonumber\\
B(f)&=&\frac{y}{f}\left[(1+f)(3f-1)+t\, (1-f)\right], \nonumber\\
C(f)&=&\frac{1}{f^2 t}\left[t\, (s+f^2)+(1-f)(2f-1)\right], \nonumber\\
D(f)&=&-\left[t+3(1-f)\right], \nonumber\\
E(f)&=&A(f) \sqrt{\frac{t}{t+1-f}},
\end{eqnarray}
where the auxiliary functions $s$, $t$, and $y$ are written
\begin{eqnarray}
\label{sty}
s(f)&=&1-2f+3f^2, \nonumber\\
t(f)&=&\sqrt{s(f)+6f^2}, \nonumber\\
y(f)^4&=&\frac{8}{s(f)+(1-f)\, t(f)}.
\end{eqnarray}
The coefficients $A$, $B$, $C$, $D$, and $E$  are monotonic functions of
$f$ and their ranges
from $f=0$ to $f=1$ are $8 \geq A(f) \geq 4$,
$0 \leq B(f) \leq 4\sqrt{2}$, $8 \geq C(f) \geq 3$,
$-4 \leq D(f) \leq -2\sqrt{2}$, and $4\sqrt{2} \geq E(f) \geq 4$.
Expression~(\ref{m2sc}) is valid for small values of
$m/\sqrt{a}$ and $\kappa$, and/or large values of $J$.

As this method relies basically on a classical approximation, it is not
possible to calculate the zero point energy of the orbital motion. Thus,
a mass formula cannot be obtained. Moreover, the dependence of the
energy as a function of the radial quantum number $\nu$ is calculated by
making a harmonic approximation around classical orbits with high values
of $J$. We cannot expect a good $\nu$ behavior for small values of the
angular momentum. A more serious flaw is that the method predicts a
linear dependence of $M^2$ as a function of $\nu$ whatever the form of
the
potential. So, we cannot be sure that the $\nu$ dependence found is the
more appropriate. A way to correct these drawbacks is to complete the
previous analysis by a BSQ method.

\subsection{BSQ method}
\label{ssec:bsq}

The basic quantities in the BSQ approach \cite{tomo62} are the action
variables,
\begin{equation}
\label{js1}
J_s=\oint p_s\, dq_s,
\end{equation}
where $s$ labels the degrees of freedom of the system, and where $q_s$
and $p_s$ are the coordinates and conjugate momenta; the integral is
performed over one cycle of the motion. The action variables are
quantized according to the prescription
\begin{equation}
\label{js2}
J_s=\nu_s+1/2,
\end{equation}
where $\nu_s$ ($\geq 0$) is an integer quantum number. This corresponds
to a WKB expansion limited to the first order in $h$ (see for instance
\cite{dunh32,bend77,robn97a,robn97b}).

The calculations for the angular momentum $J$, in the limit of great
values for this quantum number, give simply the same $J$ dependence for
$M^2$ as in expression~(\ref{m2sc}),
but with $J$ replaced by $J+1/2$. The
calculations for the radial motion are more involved. A detailed
description of the procedure is given in Ref.~\cite{brau00b} where the
case of the Hamiltonian~(\ref{hsv}) is studied for $m_1=m_2$ and $f=0$.
We use here the same technique and we expand all expressions in powers
of the meson mass $M$. Assuming that $M^2/a$ is large and $J$ finite,
and keeping only terms in $M^2$, $M$, and $1/M$, the
integral~(\ref{js1}) with Hamiltonian~(\ref{hsv}) can be written, after
tedious calculations,
\begin{equation}
\label{pi2np1}
\pi (2\nu +1) = \frac{\sqrt{Y}Z}{2a(1-2f)}-\frac{X^2}{2a(1-2f)^{3/2}}
\ln \left( \frac{Z+\sqrt{(1-2f)Y}}{X} \right),
\end{equation}
with $\nu$ the radial quantum number, and with
\begin{eqnarray}
\label{xyz}
X&=& M f + 2 m (1-f),  \nonumber \\
Y&=& M^2 -4 m^2 + 2 a \kappa (1-f) + 4 a m \kappa f /M, \nonumber \\
Z&=& M (1-f) +2 m f + a \kappa (1-2 f)/ M.
\end{eqnarray}
The expression above is valid only for $f \leq 1/2$. A similar equation
exists for $f \geq 1/2$. It is now necessary to extract $M^2$ as a
function of $\nu$ in order to obtain an analytical result usable in a
mass
formula. If we assume that quantities $m/\sqrt{a}$ and $\kappa$
are small, we can expand Eq.~(\ref{pi2np1}) in powers of these small
parameters. The first order gives ($m/\sqrt{a} = \kappa =0$)
\begin{equation}
\label{m2ep}
M^2 \approx a E'(f)(2 \nu+1) \quad \text{for} \quad \nu \gg 1,
\end{equation}
with
\begin{equation}
\label{gep}
E'(f)=2\pi \frac{1-2 f}{1-f-f^2 H(f)} \quad \text{with} \quad H(f)=
\left\{
\begin{array}{lll}
\frac{1}{\sqrt{1-2 f}}\ln \left( \frac{1+\sqrt{1-2 f}}{1-\sqrt{1-2 f}}
\right) & \text{for} & f \leq 1/2, \\
\frac{1}{\sqrt{2 f-1}} \arccos \left( \frac{1-f}{f} \right)
& \text{for} & f \geq 1/2.
\end{array}
\right.
\end{equation}
We have $E'(0)=2\pi$, $E'(1/2)=3\pi/2$, and $E'(1)=4$. This is in
agreement with results obtained in Ref.~\cite{brau00b} for the case
$f=0$.

\begin{figure}
\includegraphics*[height=8cm]{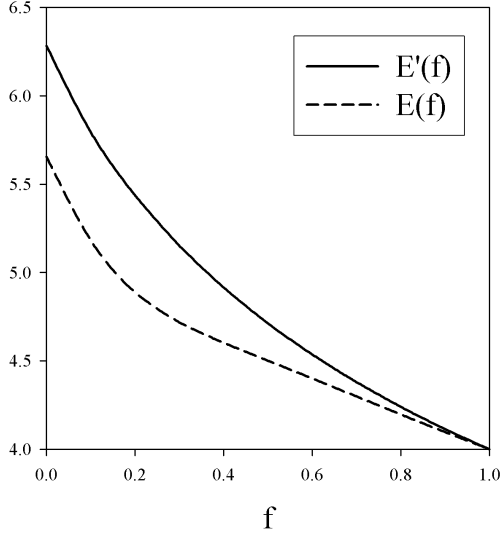}
\protect\caption{Coefficients $E(f)$ and $E'(f)$.}
\label{fig:gegep}
\end{figure}

The $\nu$ square mass dependence obtained with this method is
very similar to the one obtained with the DOS approach. But, the
coefficients $E(f)$ and $E'(f)$ are different, as it is shown on
Fig.~\ref{fig:gegep}. The approximations used to calculate these
coefficients are also very different: $E(f)$ is expected to give good
results when $J \gg \nu$, while $E'(f)$ is expected to give good results
when $\nu \gg J$.

By expanding the right hand side of Eq.~(\ref{pi2np1}) in powers of
$m/\sqrt{a}$ and $\kappa$, we obtain, at the second order,
\begin{equation}
\label{pi2np1bis}
\pi (2\nu +1) \approx \frac{M^2 \pi}{a E'(f)} +
\frac{M m f}{a (1-2 f)^{3/2}}
\left( \Delta(f) - \frac{\Delta(f) f}{\Delta(f) -f}
+2 (f-1) \ln \left[\frac{\Delta(f)}{f}-1\right] \right)
+ \kappa.
\end{equation}
where $\Delta(f)= 1 + \sqrt{1-2 f}$ (note that this expression is well
defined for $f$ in the range [0,1]). We give these expressions for the
sake of completeness. Nevertheless, as we can see below, the first order
term is sufficient for our purpose.

\subsection{Mass formula}
\label{ssec:massf}

Using results from the DOS approach and the BSQ method, we can
write a square mass formula for light mesons composed of two identical
quarks with a mass $m$ as a function of the quantum numbers $J$ and
$\nu$, and the parameters of the potentials $a$, $\kappa$, and $f$
\begin{equation}
\label{m2}
{M}^2=a\,A(f)\, (J+1/2) + B(f)\, m\, \sqrt{a\, (J+1/2)} + C(f)\, m^2 + a
\, D(f) \kappa +a\ E^{(')}(f) (2\nu +1).
\end{equation}
The coefficients $A$, $B$, $C$, $D$, and $E$ are given by
relations~(\ref{abcde})-(\ref{sty}), and the coefficient $E'$ is given
by relation~(\ref{gep}).

Actually, this formula is an approximate expression for eigenvalues of
the Hamiltonian~(\ref{hsv}). In principle, it is only valid
for large values of $J$ and/or large values of $\nu$. The coefficient
$E(f)$ will be preferred when $J \gg \nu$, while the coefficient $E'(f)$
will be chosen when $\nu \gg J$.
One can ask if the Eq.~(\ref{m2}) can give good meson mass estimations
for realistic values of the potential parameters and for small
values of the quantum numbers. In order to answer this question, and
then to test the relevance of such a formula, it is necessary to compare
the predictions of the formula with accurate numerical solutions of
Hamiltonian~(\ref{hsv}). This is the purpose of Sec.~\ref{sec:disc}.

\subsection{Addition of a constant term}
\label{ssec:const}

It is well known that it is necessary to take into account the
contribution of a constant term in potential models in order to get the
correct absolute values of the meson masses. In principle it is possible
to add a constant $\Lambda_V$ to the vector potential and a constant
$\Lambda_S$ to the scalar potential. As in the case of the confinement
potential, we can define a parameter $g$ whose value is 0 for a pure
vector constant, and 1 for a pure scalar constant
\begin{equation}
\label{csv}
\Lambda_V = (1-g) \Lambda \quad \text{and} \quad \Lambda_S = g \Lambda.
\end{equation}
A detailed discussion of the contributions of these two constants is
given in Ref.~\cite{silv98}. But it is very easy to see that the
introduction a scalar constant is equivalent to a redefinition of the
quark masses, while a vector constant simply shifts all meson masses.
Finally,
if $M(m,\Lambda)$ is the mass of a meson containing two identical quarks
interacting with the potential~(\ref{s})-(\ref{v}) supplemented by a
constant $\Lambda$ given by Eq.~(\ref{csv}), then we have
\begin{equation}
\label{mlamb}
M(m,\Lambda) = M(m+g\Lambda/2,0) + (1-g)\Lambda.
\end{equation}
Let us note that, as $\Lambda$ is always negative in realistic potential
models, we have $m+g\Lambda/2 \leq m$. Then, the formula~(\ref{mlamb})
gives a better approximation than without a constant, since it relies on
an expansion for a small mass parameter.

\subsection{Asymmetric mesons}
\label{ssec:asym}

The study of mesons containing two different quarks in the framework of
the DOS approach has been performed in Ref.~\cite{brau00a}. A formula
for the square mass very similar to the expression~(\ref{m2sc}) has been
found but with coefficients given only numerically. As these
coefficients cannot be determined analytically, they are useless for a
mass formula.

Fortunately, one can verify experimentally that the mass of a $n\bar s$
meson is in very good approximation the arithmetic mean between the
masses of the corresponding $n\bar n$ and $s\bar s$ mesons. Then, since
the mass formula~(\ref{m2}) reproduces reasonably well the mass of a
symmetric meson, as we will see in Sec.~\ref{sec:disc}, it is natural to
try the following prescription
\begin{equation}
\label{m12}
M(m_1,m_2) = \frac{M(m_1,m_1) + M(m_2,m_2)}{2},
\end{equation}
where $M(m,m)$ is the mass of a symmetric meson given by
Eq.~(\ref{m2}), and $M(m_1,m_2)$ the mass of an asymmetric meson
containing two quarks with masses $m_1$ and $m_2$. Equation~(\ref{m12})
is then purely phenomenological.

It is worth noting that the scalar potential and the scalar constant are
equally shared between the two quarks in a symmetric meson. This is
not the case for an asymmetric meson. The relation~(\ref{m12}) does
not take into account this situation. We will verify below that this
does not spoil the quality of the mass formula for asymmetric mesons.

\section{Discussion of the model}
\label{sec:disc}

In order to test the relevance of the square mass formula given by
Eqs.~(\ref{m2}), (\ref{mlamb}), and (\ref{m12}), we have compared its
predictions with the exact eigenvalues of Hamiltonian~(\ref{hsv}). They
are obtained with a great accuracy by using a Lagrange-mesh calculation
method for semirelativistic equations \cite{sema01}. This method is
described in the appendix, with the modification which must be made in
order to handle the spinless Salpeter equation with mixed scalar-vector
potentials.

\begin{figure}
\includegraphics*[height=8cm]{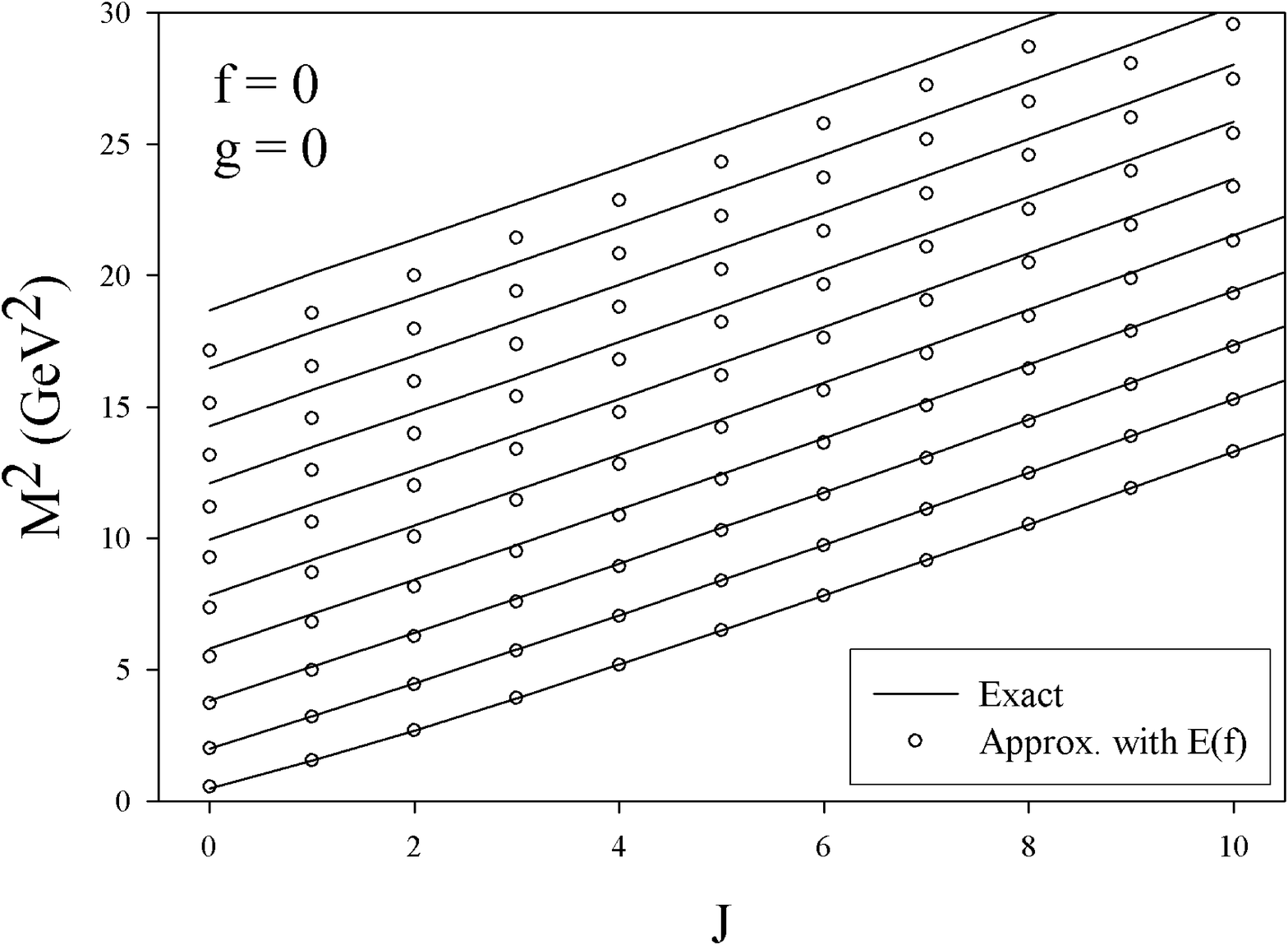}
\protect\caption{Square masses $M^2$ of $n\bar n$ mesons as a function
of $\nu$ and $J$ quantum numbers, with parameters of
Ref.~\protect\cite{fulc94},
and for $g=0$ and $f=0$. Solid lines join exact solutions of the
spinless Salpeter equation. Approximate results from Eq.~(\ref{m2}) with
the coefficient $E(f)$ are indicated by circles.}
\label{fig:f0e}
\end{figure}

\begin{figure}
\includegraphics*[height=8cm]{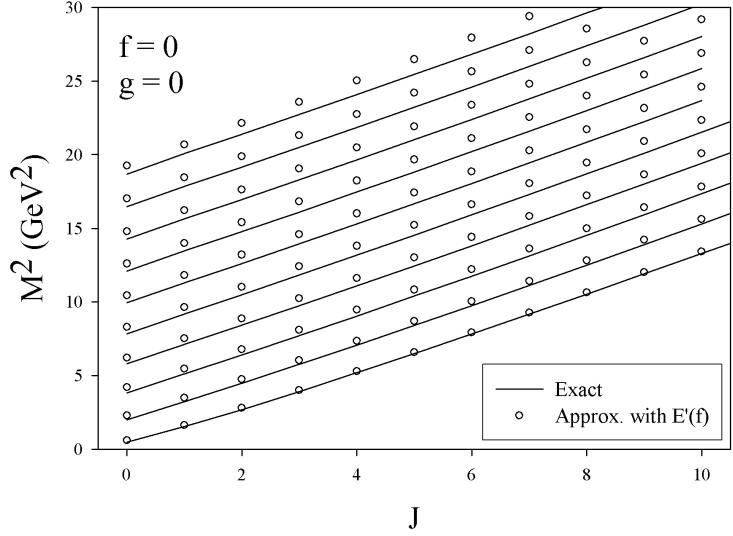}
\protect\caption{Same as Fig.~\ref{fig:f0e} but for
Eq.~(\ref{m2}) with the coefficient $E'(f)$.}
\label{fig:f0ep}
\end{figure}

\begin{figure}
\includegraphics*[height=8cm]{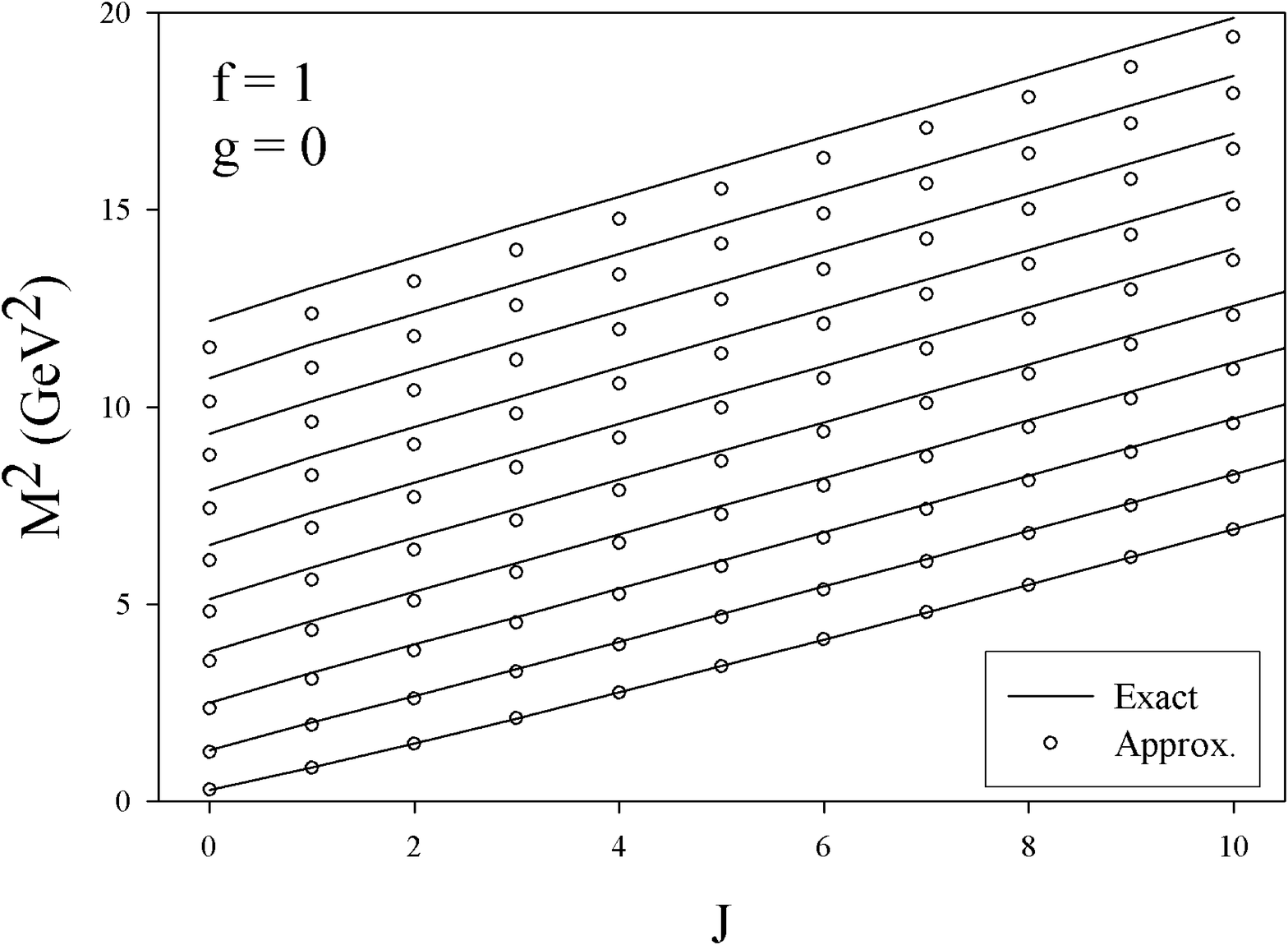}
\protect\caption{Square masses $M^2$ of $n\bar n$ mesons as a function
of $\nu$ and $J$ quantum numbers, with parameters of
Ref.~\protect\cite{fulc94},
and for $g=0$ and $f=1$. Solid lines join exact solutions of the
spinless Salpeter equation. Approximate results from Eq.~(\ref{m2}) are
indicated by circles ($E'(1)=E(1)$).}
\label{fig:f1eep}
\end{figure}

\begin{figure}
\includegraphics*[height=8cm]{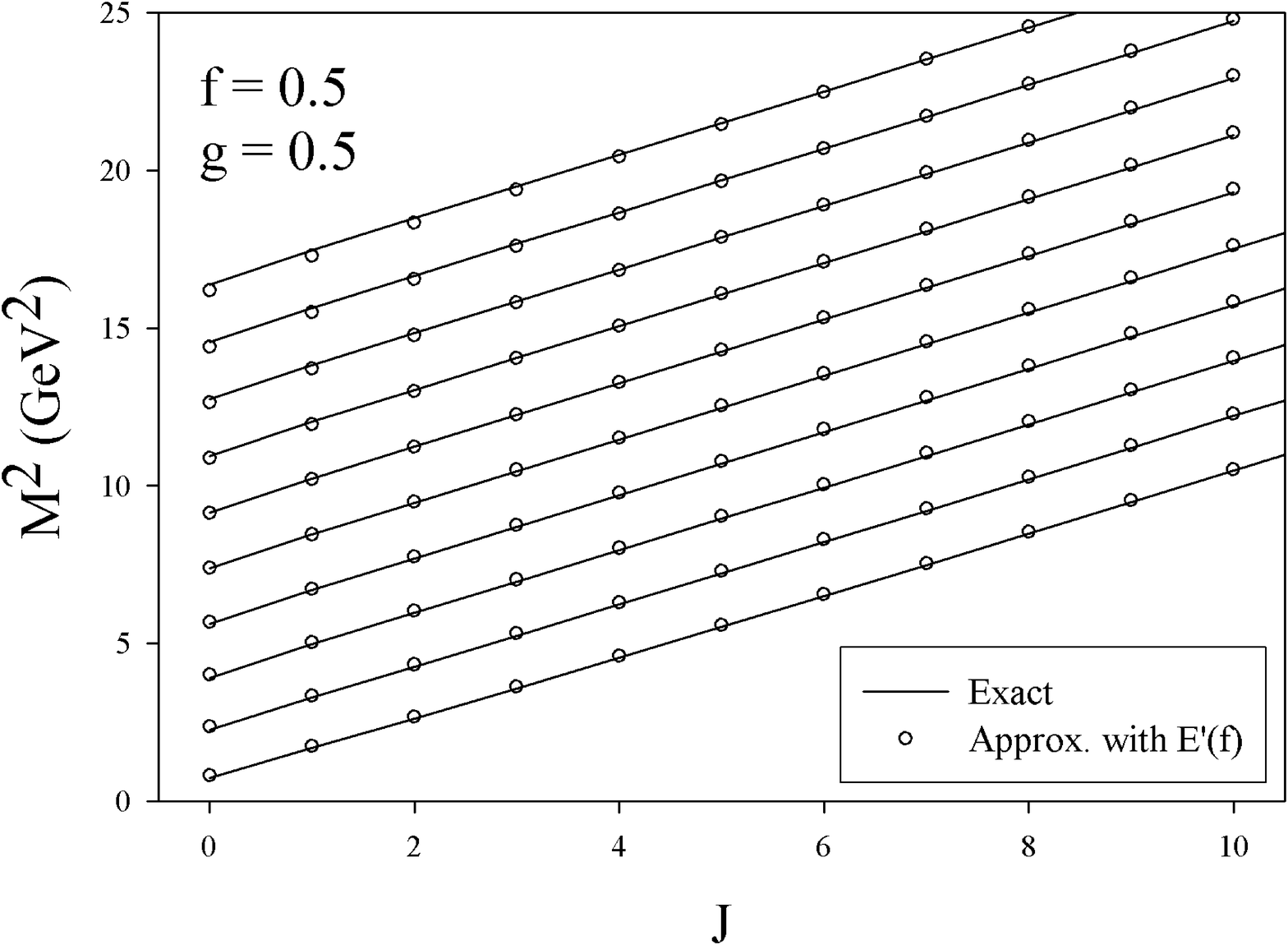}
\protect\caption{Square masses $M^2$ of $n\bar s$ mesons as a function
of $\nu$ and $J$ quantum numbers, with parameters of
Ref.~\protect\cite{fulc94},
and for $g=0.5$ and $f=0.5$. Solid lines join exact solutions of the
spinless Salpeter equation. Approximate results from Eq.~(\ref{m12})
with the coefficient $E'(f)$ are indicated by circles.}
\label{fig:ns}
\end{figure}

Realistic values for the quark masses and parameters of the potential
are taken from a simple semirelativistic model described in
Ref.~\cite{fulc94}: $m_n=0.150$ GeV, $m_s=0.364$ GeV, $a=0.203$ GeV$^2$,
and $\kappa=0.437$ ($m_n/\sqrt{a}=0.33$ and $m_s/\sqrt{a}=0.81$). In
this paper, several constants are used, depending
on the quark contents of the meson. Here we have just used the constant
for $n\bar n$ meson, $\Lambda=-0.599$ GeV. The parameters are chosen to
reproduce the main features of the meson spectra with $f=g=0$. As,
We have used these parameters in our square mass formula with varying
values of $f$ and $g$, we cannot obtain good spectra for all values of
these two quantities. But the purpose of this work is simply to show the
relevance of the mass formula. In Fig.~\ref{fig:f0e} to
Fig.~\ref{fig:ns},
we compare the exact eigenvalues of Hamiltonian~(\ref{hsv}) with the
predictions of the mass formula given by Eqs.~(\ref{m2}), (\ref{mlamb}),
and (\ref{m12}).

In Fig.~\ref{fig:f0e}, the square $n\bar n$ meson masses are plotted as
a
function of $J$ and $\nu$ for parameter values $f=g=0$, the coefficient
$E(f)$ being used ($E(f)$ and $E'(f)$ are the most different for $f=0$).
We can see that for small values of $\nu$ or large values of $J$, the
exact and approximate results are in good agreement. This situation is
expected when the coefficient of the quantum number $\nu$ is obtained
with the DOS approach. For instance, for $J=0$, the relative error on
square mass increases regularly from 3.3\% at $\nu=2$ to 8.3\% at
$\nu=9$ (there are irregularities between $\nu=0$ and $\nu=2$). In
Fig.~\ref{fig:f0ep}, the same quantities are plotted, but the
approximate
results are calculated with the coefficient $E'(f)$. In this case, for
$J=0$, the absolute error on square mass increases regularly from 0.120
GeV at $\nu=0$ to 0.570 GeV at $\nu=9$, but in the same time, the
relative error decreases from 24.5\% to 3.1\%. This is again the
expected behavior for a coefficient of the quantum number $\nu$ obtained
with the BSQ method. In the sector where the coefficients $E(f)$ and
$E'(f)$ are different, it is interesting to choose the coefficient of
the quantum number $\nu$ following the values of the quantum numbers $J$
and $\nu$. With this constraint, the results of the square mass formula
are quite good.

When $f=1$, the situation is simpler since $E(f)$ and $E'(f)$ are
identical. In Fig.~\ref{fig:f1eep}, the square $n\bar n$ meson masses
are
plotted as a function of $J$ and $\nu$ for parameter values $f=1$ and
$g=0$. Again the agreement between exact and approximate results is
good. The qualities of the mass formula seems to deteriorate with
increasing values of $\nu$. Actually, for $J=0$, the absolute error on
square mass increases regularly from 0.148 GeV at $\nu=2$ to 0.674 GeV
at $\nu=9$, but in the same time, the relative error decreases slowly
from 5.9\% to 5.5\% (there are irregularities between $\nu=0$ and
$\nu=2$).

Finally, we have tested the formula~(\ref{m12}) by calculating the
square $n\bar s$ meson masses as a function of $J$ and $\nu$ for
parameter values $f=0.5$ and $g=0.5$ with the coefficient $E'(f)$.
Results are plotted in Fig.~\ref{fig:ns}. We can see that the
prescription works very well.

\section{Concluding remarks}
\label{sec:concl}

The main features of the spectra of light mesons, except pseudoscalar
ones, can be reproduced with a spinless Salpeter equation supplemented
with the Cornell interaction \cite{fulc94,brau98}. We have shown that
the eigenvalues of this simple Hamiltonian can be obtained, within a few
percents of relative error, by a mass formula. This relation gives the
square mass of light mesons as a function of quantum numbers $J$ and
$\nu$ and the parameters of the potential. One could expect that the
formula is only valid for high values of these quantum numbers and/or
for very small values of parameters $m/\sqrt{a}$ and $\kappa$. Outside
these limits, we have remarked that the formula only differs slightly
from exact results for physical values of parameters. This allows to
apply our approximation for physical situations and to compare our
calculations with experiment.

The values of the string tension $a$ and of the strength $\kappa$ of the
Coulomb-like potential can, in principle, be computed from lattice
calculations. The values of the constituent masses and of the constant
potential are more difficult to obtain, as well as the quantities $f$
and $g$. These parameters, in particular the scalar-vector mixture in
the confinement, could be determined by a fit of our square mass formula
on available experimental data. Unfortunately, some uncertainties on
experimental meson masses, leading to large intervals of possible values
for the parameters, make difficult such a determination for the moment
\cite{silv98}.

The meson mass formula obtained here shares some similarities with other
formulae, but it relies on different basis: Instead of relying on
a spectrum generating algebra \cite{iach91} or on a completely
phenomenological point of view \cite{sema95}, it is assumed here that a
semirelativistic potential model allows a good description of the main
features of meson spectra, as it is the case in Ref.~\cite{cea82}. The
meson mass dependence on some usual parameters of semirelativistic
potential model is then obtained with a good approximation.

In our study, we completely neglect the quark spin. In usual models the
spin-dependent part of the potential is the hyperfine interaction
stemming from the one-gluon exchange interaction (and may be from the
vector part of the confinement) \cite{luch91}. In other models, this
spin-dependent part stems from an instanton induced interaction
\cite{brau00a,blas90}. Except for pseudoscalar mesons, in both cases,
the contribution of the spin to the meson masses is small or vanishing
with respect to orbital or vibrational excitations. Then, despite the
absence of spin dependence in our Hamiltonian, our mass formula can
describe a large sample of mesons.



\appendix

\section*{Numerical method}

The semirelativistic Lagrange-mesh method can be used to compute the
eigenvalues and the eigenfunctions of the following
spinless Salpeter Hamiltonian
\begin{equation}
\label{h_sr}
H=\sqrt{\vec p\,^2 + m_1^2} + \sqrt{\vec p\,^2 + m_2^2} + V(r).
\end{equation}
A detailed presentation of the technique is given in Ref.~\cite{sema01}.
Only the main points of the method are given here. A variational
calculation is performed with the trial state
\begin{equation}
\label{psi}
|\psi \rangle = \sum_{j=1}^N C_j |f_j \rangle \quad \text{where} \quad
\langle \vec r\,|f_j \rangle = \frac{f_j (r/h)}{\sqrt{h}r} Y_{\ell m}
(\hat r).
\end{equation}
The coefficients $C_j$ are linear variational parameters and the scale
factor $h$ is a non-linear parameter aimed at adjusting the mesh to the
domain of physical interest. The functions $f_j (x)$ are such that
\begin{equation}
\label{fjxi}
f_j (x_i) = \lambda_i^{-1/2} \delta_{ij}.
\end{equation}
The numbers $x_i$, which are the zeros of a Laguerre polynomial of
degree $N$, and the numbers $\lambda_i$ are connected with a Gauss
quadrature formula
\begin{equation}
\label{int}
\int^{\infty}_0 g(x) dx \approx \sum_{k=1}^{N} \lambda_k g(x_k).
\end{equation}

\par At the Gauss approximation,
$\langle f_i|f_j\rangle \approx \delta_{ij}$, the
potential matrix elements are simply given by
\begin{equation}
\label{vij}
\langle f_i|V(r)|f_j\rangle \approx V(h x_i)\,\delta_{ij}.
\end{equation}
The computation of the matrix elements
$\langle f_i|\sqrt{\vec p\,^2 + m^2}|f_j\rangle$ is obtained from the
calculation of the matrix elements
$\langle f_i|\vec p\,^2 + m^2|f_j\rangle$. These last quantities can be
easily obtained at the Gauss approximation with analytical formulae
\cite{sema01}.
Despite the use of an approximate quadrature rule, a very high accuracy
can be attained with a small number of basis states.

\par When a Hamiltonian of the form~(\ref{hsv}) is considered, the
matrix elements of the operators
$\sqrt{\vec {p\,}^2+\left( m_k+\alpha_k S(r)\right)^2}$ must be
computed. Again, they can be obtained from the matrix elements of the
square of the operators. From Eq.~(\ref{vij}), we have
\begin{equation}
\langle f_i|\vec {p\,}^2+\left( m_k+\alpha_k S(r) \right)^2|f_j\rangle
\approx \langle f_i|\vec p\,^2 + m_k^2|f_j\rangle + \left( 2m_k\alpha_k
S(h x_i) + \alpha_k^2 S(h x_i)^2 \right)\delta_{ij}.
\label{tsv}
\end{equation}
We have checked that this procedure allows the computation of the
eigenvalues and the eigenfunctions of the spinless Salpeter
Hamiltonian~(\ref{hsv}) with a high accuracy.

\clearpage

\end{document}